\pdfoutput=1
\documentclass[twocolumn]{aastex631} 
\pdfoutput=1
\shorttitle{Terminator Habitability}
\shortauthors{Lobo et al.}
\graphicspath{{./}{figures/}}

\usepackage[colorinlistoftodos]{todonotes} 
\usepackage{gensymb}

\begin{document}

\title{Terminator Habitability: the Case for Limited Water Availability on M-dwarf Planets}

\author[0000-0003-3862-1817]{Ana H. Lobo}
\affiliation{University of California, Irvine \\
Department of Physics \& Astronomy \\
4129 Frederick Reines Hall \\
Irvine, CA 92697-4575, USA}

\author[0000-0002-7086-9516]{Aomawa L. Shields}
\affiliation{University of California, Irvine \\
Department of Physics \& Astronomy \\
4129 Frederick Reines Hall \\
Irvine, CA 92697-4575, USA}
\affiliation{NASA NExSS Virtual Planetary Laboratory, Seattle, WA}

\author{Igor Z. Palubski}
\affiliation{University of California, Irvine \\
Department of Physics \& Astronomy \\
4129 Frederick Reines Hall \\
Irvine, CA 92697-4575, USA}

\author[0000-0002-7188-1648]{Eric Wolf}
\affiliation{University of Colorado, Boulder \\
Laboratory for Atmospheric and Space Physics, \\
 Department of Atmospheric and Oceanic Sciences, \\
 Boulder, CO} 
\affiliation{NASA NExSS Virtual Planetary Laboratory, Seattle, WA}
\affiliation{NASA GSFC Sellers Exoplanet Environments Collaboration, Greenbelt, MD}



\begin{abstract}

Rocky planets orbiting M-dwarf stars are among the most promising and abundant astronomical targets for detecting habitable climates. Planets in the M-dwarf habitable zone are likely synchronously rotating, such that we expect significant day-night temperature differences, and potentially limited fractional habitability. Previous studies have focused on scenarios where fractional habitability is confined to the substellar or ``eye" region, but in this paper we explore the possibility of planets with terminator habitability, defined by the existence of a habitable band at the transition between a scorching dayside and a glacial nightside. Using a global climate model, we show that for water-limited planets it is possible to have scorching temperatures in the ``eye" and freezing temperatures on the nightside, while maintaining a temperate climate in the terminator region, due to a reduced atmospheric energy transport. Whereas on water-rich planets, increasing stellar flux leads to increased atmospheric energy transport and a reduction in day-night temperature differences, such that the terminator does not remain habitable once the dayside temperatures approach runaway or moist greenhouse limits. We also show that, while water-abundant simulations may result in larger fractional habitability, they are vulnerable to water loss through cold-trapping on the nightside surface or atmospheric water vapor escape, suggesting that even if planets were formed with abundant water, their climates could become water-limited and subject to terminator habitability.

\end{abstract}

\keywords{planets and satellites: terrestrial planets --- planets and satellites: atmospheres}


\section{Introduction} \label{sec:intro}

Rocky planets in the habitable zones of M-dwarf stars are currently among the most promising astronomical targets for planetary climate characterization studies and the detection of habitable surface climate. M dwarfs, which make up $\sim$70\% of all stars \citep{Bochanski2010}, are expected to have abundant rocky Earth-sized planets \citep{Mulders2015}. Also, their reduced luminosity results in a habitable zone that is only $\sim$0.2 au from the star \citep{Kasting1993, Kopparapu2013}, which facilitates detection \citep{Nutzman2008}. While M-dwarf prospects for habitability have been extensively debated \citep{Scalo2007,Tarter2007,Shields2016}, over the course of the last decade many of the most promising candidates for habitability have been discovered orbiting M dwarfs, such as TRAPPIST-1 \citep{Gillon2017}, Proxima Centauri \citep{Anglada2016}, and TOI-700 \citep{Gilbert2020}. With the recent launch and deployment of the James Webb Space Telescope, it may soon be possible to characterize the atmospheres of terrestrial planets in the habitable zones of M dwarfs \citep{Kreidberg2016, Morley2017, Fauchez2019}. In preparation for upcoming observations, it is increasingly important that we understand the full range of possible M-dwarf planetary climates and their prospects for habitability.   

The habitable zone's close proximity to the star implies that the planets are subject to rapid tidal locking, and are likely to be synchronously rotating such that they have a permanent dayside \citep{Armitage2009, Barnes2017}. Without incoming radiation, the nightside climate is determined by atmospheric and ocean energy transport, and could be subject to atmospheric collapse \citep{Joshi1997}. Whereas, on the highly irradiated dayside, greenhouse effects could be amplified with increased atmospheric water vapor content \citep{Ingersoll1969}, potentially leading to a runaway state. Despite the nightside's inevitable top-of-atmosphere radiative deficit, if the day-to-night energy transport is sufficiently intense, it is possible for a large portions of a synchronously rotating planet to remain above freezing \citep[e.g.][]{Hu2014, delgenio2019, Yang2019oceandyn}, without entering a runaway greenhouse state. Even neglecting ocean transport, aquaplanet simulations can achieve moderate temperatures on synchronously rotating planets \citep[e.g.][]{Joshi2003, Merlis2010, Edson2011}. A temperate climate is achievable even when accounting for the longer wavelength stellar spectra of M dwarfs, which results in changes to key radiative feedbacks \citep[e.g.][]{Shields2016, Kopparapu2016, Turbet2016, Komacek2019}. Yet, while we can easily simulate an arbitrary habitable M-dwarf planet, the range of habitable planetary configurations is by no means fully constrained, especially when we consider both ocean-covered and land planets. 

Water is a minimum requirement for life as we know it. Therefore, we are naturally most interested in the climate of water worlds. Note our Earth is effectively a water world even with $\sim$30 percent of its surface covered by continents. For simplicity, in exoplanet modeling studies these are often assumed to be completely ocean-covered worlds, but that is not necessarily the case, especially for M-dwarf planets. Determining the amount of water expected on a rocky M-dwarf planet in the habitable zone is no simple matter. Among other things, it depends on whether we expect the planets to have formed in-situ or have migrated. Due to the low luminosity of M dwarfs, their ice lines are relatively close to the habitable zone ($\sim$ 0.3 AU), such that migration in the inner disk could facilitate water-rich habitable zone planets \citep{Ogihara2009}. On the other hand, in situ formation of volatile rich planets would be less likely \citep{Lissauer2007, Raymond2007, Raymond2022}, even when accounting for volatile delivery through the migration of icy planetesimals \citep{Ciesla2015}. There is also a higher risk of water loss for M-dwarf planets due to early periods of increased stellar flux and energetic flare activity \citep{Bolmont2017, Tian2015, Luger2015, Ramirez2014}, as well as periods of intense tidal heating prior to reaching spin-orbit resonance \citep{Barnes2013}. For the purpose of this study, we are specifically interested in moist M-dwarf planets, but it is possible that these could be predominantly water-limited land planets rather than ocean-covered worlds. Therefore, our study compares both both water-rich ocean-covered scenarios, and water-limited land planet scenarios.

\begin{figure}[h]
\centering
\includegraphics[width=.48\textwidth]{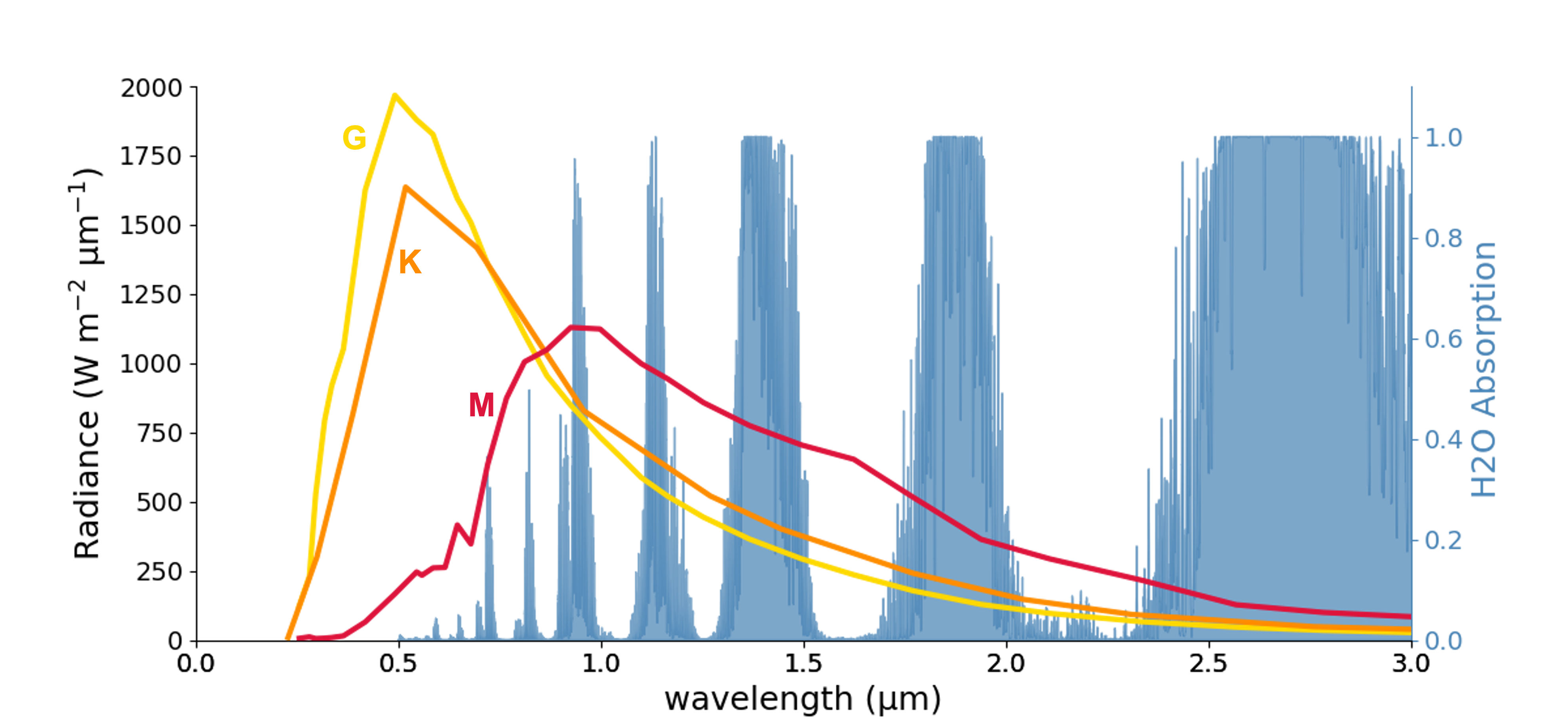}
\caption{\label{fig:sw_profiles} Stellar spectra for AD Leonis (red), HD22049, a K-star (orange), and the Sun (yellow). Blue lines show water absorption values using the Lorentz profile, from the HITRAN Database. }
\end{figure}

The presence of water on the surface of a planet can have a wide range of effects on the climate, for example, altering the surface heat capacity \citep{Cronin2013,Donohoe2014}. Water also impacts the surface albedo, whether due to low albedo liquid water or the ice-albedo feedback. 
Water plays a key role in the radiative budget, influencing cloud formation and structure \citep{Stevens2005}, which in turn affects planetary albedo \citep{Donohoe2011} and greenhouse effect intensity. It can also alter the atmospheric energy transport \citep{Held2006}, enhance local energy storage \citep{Donohoe2014, Lobo2020}, and functions as an important greenhouse gas \citep{Held2000} with a strong positive climate feedback. Due to water's various climate feedbacks and its effect on the atmospheric structure, the habitable zone of a water-limited Earth twin is broader than that of an aquaplanet Earth \citep{Abe2011}.
But, while water's impact on climate is well-understood for Earth, many of these fundamental climate feedbacks behave differently on M-dwarf planets due to the lower frequency of the stellar radiation. 

We typically refer to the stellar radiation as the ``shortwave" flux (abbreviated SW), to distinguish it from a planet's ``longwave" thermal emissions (LW). However, compared to other stars M-dwarfs have relatively long wavelength emissions. M dwarf effective temperatures range from 2000 to 3800K, such that their peak emissions are at near-IR wavelength (Fig.~\ref{fig:sw_profiles}) and a large fraction of the SW overlaps with absorption bands for CO$_2$ and H$_2$O \citep{Kasting1993, Selsis2007}.  Water's absorption is particularly important, with multiple absorption bands near the spectra peak. For reference, we plot H$_2$O absorption in Fig.~\ref{fig:sw_profiles}, using values from the \textit{HITRAN} database.  
This implies that unlike Earth's atmosphere, which is predominantly heated from below due to SW absorption at the surface, a moist M-dwarf planetary atmosphere can be heated at various levels. It is also important to note that ice and snow albedos are also lower for IR radiation \citep{Dunkle1956}, such that the ice-albedo feedback is weaker. On M-dwarf planets the effects of longer-wavelength radiation result in warmer climates than would be obtained for equal stellar flux from a higher frequency emitter, such as G and F dwarfs \citep{Shields2013, Shields2014, Shields2016}. Therefore, the climate response to increased stellar radiation on M-dwarf planets is not necessarily similar to Earth's, and climate scenarios must be explored in the context of specific stellar types.

In this paper, we explore climate at the inner edge of the M-dwarf habitable zone, to determine how fractional habitability changes as dayside temperatures start to exceed habitable limits. In particular, we seek to determine whether it is possible to sustain a habitable band at the transition between a scorching dayside and a glacial nightside, a scenario which we refer to as terminator habitability, or whether fractional habitability becomes impossible once the substellar region surface temperatures surpass habitable limits. While there is no broad consensus in terms of a habitable temperature range, here we use a relatively narrow definition of temperatures between  0\textdegree C and 50\textdegree C, to provide a more conservative measure and to better track how fractional habitability changes \citep{Shields2016}. Therefore, we are exploring whether planets with a medium or large day-to-night temperature gradient can exist specifically near the inner edge of the habitable zone, such that the dayside temperatures surpass 50\textdegree C, and the resultant consequences in terms of climate and fractional habitability.

For this study, we consider Earth-like planets orbiting AD Leonis, which has a spectral classification of M3.5V, such that synchronously rotating planets in its habitable zone are in a slowly rotating regime \citep{HaqqMisra2018, Noda2017}. On these planets, the rossby deformation radius is smaller than the planetary radius, resulting in a dynamical regime with strong convective activity at the substellar point, a thermally direct day-to-night circulation, and tend to have roughly symmetric terminator properties. For dimmer M-dwarf stars, the habitable zone can include synchronously rotating planets that are more rapidly rotating, which have distinct dynamical properties, including a stronger mean zonal flow that produces a band-like structure at the equator \citep[e.g.][]{HaqqMisra2018}. But here we focus on dynamics relevant to the habitable zone of brighter M dwarfs, which represent the majority of observed M-dwarf stars \citep{HIPPARCOS1997} and host the majority of exoplanets detected so far (according to the \textit{NASA Exoplanet Archive}). 

We use a 3D global climate model to determine whether it is possible to sustain a temperature gradient large enough for a terminator habitability scenario, and explore the implications of terminator habitability for future climate characterization studies. It is not our goal to precisely quantify the habitable zone edge, given that its location is dependent on a large range of properties, including planetary radius and surface gravity \citep{Yang2019radius, Thomson2019}, among many other factors \citep[see e.g.][]{Meadows2018, Kasting1993, Kopparapu2013, Yang2013, Yang2014HZRotation}, but rather to explore the mechanisms through which the atmosphere responds to increased stellar flux, including changes in the radiative budget and atmospheric energy transport, to determine the viability of these surface climate configurations. We begin with a study of water-abundant aquaplanet simulations (Sections \ref{subsec:Water Vapor}), followed by a comparison with water-limited land planet simulations (Section \ref{sec:water limited}). Note that for simplicity, we will define habitability based only on surface temperature for the majority of the paper, but we discuss long term water availability and the implications for habitability in Section \ref{sec:long term} and \ref{sec:water_avail}.

\section{Methods} \label{sec:Methods}

In this study, we use ExoCAM \citep{githubExoCAM, Wolf2022}, which is a modified version of the Community Atmosphere Model (CAM4), with correlated-k radiative transfer (ExoRT)\footnote{
During revisions, a minor interpolation error was found and fixed in the ExoCAM ExoRT package \citep[commit 0223865,][]{githubExoCAM}. We ran simulations, branched from the previous spun-up cases, using the updated model. These confirmed that the bug did not impact our results. 
}, and a finite-volume dynamical core. A complete description of the code and its lineage is available in \citet{Wolf2022} and references within. In order to isolate the effects of instellation on the day-night temperature contrasts and terminator habitability thresholds, our simulations are Earth-like in terms of radius and gravity, with an atmospheric composition of 40Pa CO$_2$, 0.17Pa CH$_4$, and the remainder N$_2$, totaling 1 bar for the dry component of the atmosphere, with variable H$_2$O also included. Simulations were run with a horizontal resolution of $4\degree \times 5\degree$, and 40 vertical levels. We use the stellar spectra of AD Leonis \citep{Reid1997, Segura2005}, shown in Fig.~\ref{fig:sw_profiles}, prescribing zero eccentricity and zero obliquity, such that there is no seasonal cycle. We also calculate the two-band snow and ice albedos, which we refer to as the visible ($0.25 < \lambda < 0.75 \mu m$) and near-IR ($0.75 < \lambda < 2.5  \mu m$), weighted by the stellar spectrum:

\begin{deluxetable}{ccc}[h]
\tablenum{1}
\tablecaption{Albedo Values}
\tablewidth{0pt}
\tablehead{
\colhead{} & \colhead{Visible} & \colhead{Near-IR} 
}
\startdata
Snow    & 0.97 & 0.48\\ 
Ice   & 0.64 & 0.17  
\enddata
\end{deluxetable}

All simulations use a time step of 1800s and are run until they reach steady state. Results are shown as a time average of the last 10 years. Two simulations did not reach steady state (Aq25 and Aq25h), and instead failed due to numeric instabilities as they entered a runway greenhouse state. Results for Aq25 are briefly discussed using the average of the last stable month of model time. Results are shown on a planetary grid where the substellar point is at 0\degree \space longitude and 0\degree \space latitude. Figures with a vertical profile use the CESM hybrid sigma-pressure coordinate, plotted on a scale of 0-1 where 1 is the simulation's reference surface pressure (P$_o$).

The aquaplanet simulations (Aq) have uniform 50m mixed layer depth and no ocean energy transport. The Aq34 aquaplanet is located 0.15au from the star. We use this planet as a starting point for our analysis because, for a planet orbiting AD Leonis, this orbital distance results in an Earth-like solar constant and a mostly temperate dayside climate. To explore the inner edge of the habitable zone and the transition into runaway greenhouse states, we also use simulations with smaller orbital radii. Per Kepler's third law, a reduction in orbital radii is accompanied by a change in orbital period, such that 

\begin{equation}
    P= \Big( a^3 \frac{M_{\bigodot}}{M_{\bigstar}} \Big)^{\frac{1}{2}},
\end{equation}
where $P$ is the orbital period in years, $M_{\bigstar}$ is stellar mass, $M_{\bigodot}$ is the Sun's mass, and $a$ is the orbital radius in au. Therefore, to reduce orbital radii, and increase stellar flux, we compare simulations with orbital and rotational periods of 34, 30, and 25 days (Table~\ref{tab:data_list}). We also include aquaplanet simulations with atmospheric surface pressure reduced to 0.5bar (Aqh).

Given our interest in the greenhouse effect, and the enhanced importance of water vapor as an absorber for redder stars, we also use land planet simulations (L) where the planet's water is limited. These simulations are initialized without soil or ground water, with the ground properties of sand everywhere (no oceans). All water is initially homogeneously distributed in the atmosphere, where we prescribe the specific humidity (Q). This configuration uses the Community Land Model (CLM) \citep{NCARTN}, such that over time, water can remain in the atmosphere, precipitate, accumulate in the soil, or solidify as snow/ice. We emphasize that these water-limited simulations are significantly different from the aquaplanet simulations, because aquaplanet simulations have unlimited water available for evaporation at each grid box not covered in ice.

\begin{deluxetable}{cccllc}[h]
\tablenum{2}
\tablecaption{List of Simulations \label{tab:data_list}}
\tablewidth{0pt}
\tablehead{
\colhead{Name} & \colhead{Type} & \colhead{$P_{\text{orb}}$ (days)} & \colhead{a (au)} &
\colhead{F (W/m$^2$) } & \colhead{P$_o$}
}
\startdata
Aq34  & Aquaplanet & 34 days & 0.154 & 1378 & 1.0 bar \\ 
Aq34h & Aquaplanet & 34 days & 0.154 & 1378 & 0.5 bar \\ 
Aq30  & Aquaplanet & 30 days & 0.141 & 1629 & 1.0 bar \\ 
Aq30h & Aquaplanet & 30 days & 0.141 & 1629 & 0.5 bar \\ 
Aq25  & Aquaplanet & 25 days & 0.125 & 2077 & 1.0 bar \\ 
Aq25h  & Aquaplanet & 25 days & 0.125 & 2077 & 0.5 bar \\ 
L34Qe3  & Q=0.0010 & 34 days & 0.154 & 1378 & 1.0 bar \\ 
L34Qe4  & Q=0.0001 & 34 days & 0.154 & 1378 & 1.0 bar \\ 
L25Qe4 & Q=0.0001 & 25 days & 0.125 & 2077 & 1.0 bar \\ 
\enddata
\end{deluxetable}

\section{Results} \label{sec:Results}

\subsection{Water-Abundant Worlds} \label{subsec:Water Vapor}

\begin{figure}[h]
\vspace{-3mm}
\centering
\includegraphics[width=0.47\textwidth]{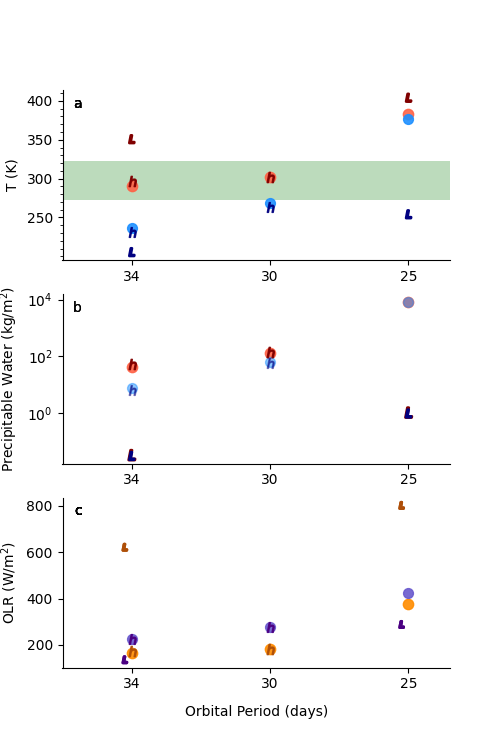}
\caption{\label{fig:min_max_T} Minimum (blue) and maximum (red) surface temperature (a) and precipitable water (b) for 1bar (circle) and 0.5bar (h) simulations. L markers show  values for  L34Qe4 and L25Qe4 simulations. The shaded region highlights the temperature range between 0 and 50\degree C, where we might expect climate favorable to complex life forms. Subplot c shows substellar (orange) and antistellar (purple) TOA OLR values.}   
\end{figure}

\begin{figure}[h]
\centering
\includegraphics[width=0.5\textwidth]{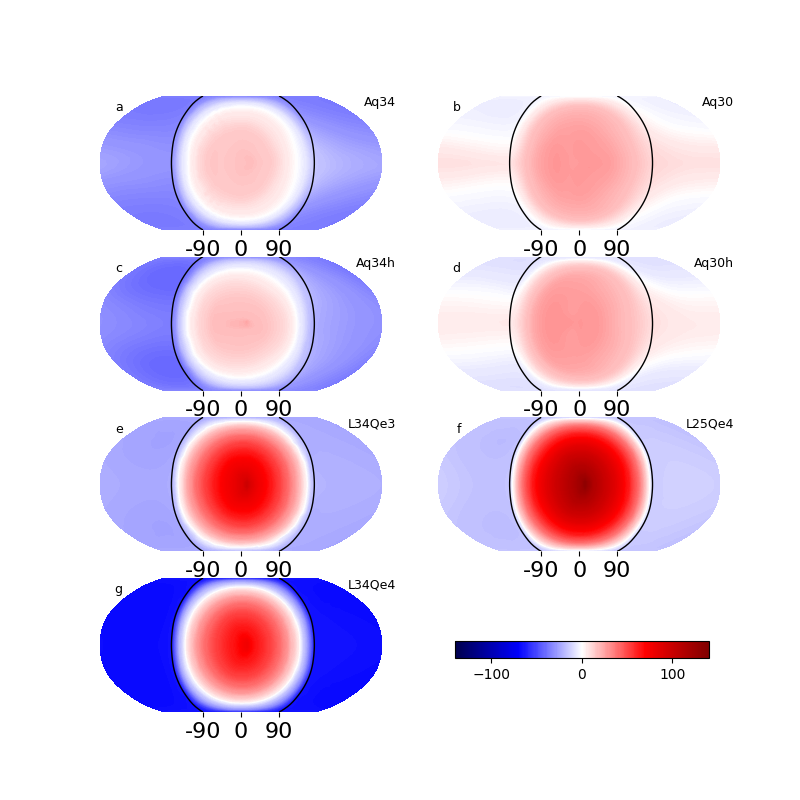}
\caption{\label{fig:TS_robinson} Surface temperatures ($\degree$C), plotted with substellar point at center. Black lines indicate the terminator. }
\end{figure}

We begin with a simple comparison of the water-abundant cases (Aq34, Aq30, and Aq25), examining the maximum and minimum surface temperatures. By definition, in order for terminator habitability to occur a planet must sustain large day-night temperature gradients. However, the simulations show that there is a significant reduction in the surface temperature range as the planet moves closer to the star (Fig.~\ref{fig:min_max_T}a) in agreement with the findings of \citet{Yang2019oceandyn}, \citet{HaqqMisra2018}, and \citet{Noda2017}. Comparing Aq34 and Aq30, we note that Aq30 is on average 25K warmer, but the day-night temperature contrast ($\Delta$T) is only 34K (Table~\ref{tab:mean_values}). This is unfavorable for terminator habitability because, as the planet's mean temperature increases, the nightside warms more than the dayside, and the day-night contrast becomes small. Therefore, at the inner edge of the habitable zone, we would expect planets to either be mostly habitable (e.g. Fig.~\ref{fig:TS_robinson}b) or, if their temperatures become sufficiently high, entirely uninhabitable due to a global runaway greenhouse effect. In this section, we explore the mechanisms that lead to small day-night temperature differences in the water-abundant planets. 

\begin{deluxetable}{cccllc}[h]
\tablenum{3}
\tablecaption{Planetary surface temperatures and fractional habitability (FH).  \label{tab:mean_values}}
\tablewidth{0pt}
\tablehead{
\colhead{ } & \colhead{Mean T} & \colhead{T$_{\text{max}}$} & \colhead{T$_{\text{min}}$} &
\colhead{$\Delta$T} & \colhead{FH 0 - 50\degree C}
}
\startdata
Aq34    &   259K &   291K &   236K &   54K  &   32\%    \\
Aq34h   &   257K &   297K &   232K &   66K  &   30\%    \\
Aq30    &   283K &   301K &   268K &   33K  &   79\%    \\
Aq30h   &   282K &   302K &   264K &   38K  &   75\%    \\
L34Qe3  &   279K &   370K &   248K &   122K &   24\%    \\
L34Qe4  &   247K &   350K &   205K &   145K &   24\%    \\
L25Qe4  &   295K &   404K &   254K &   150K &   16\%    \\
\enddata
\end{deluxetable}

\begin{figure}[h]
\centering
\includegraphics[width=0.3\textwidth]{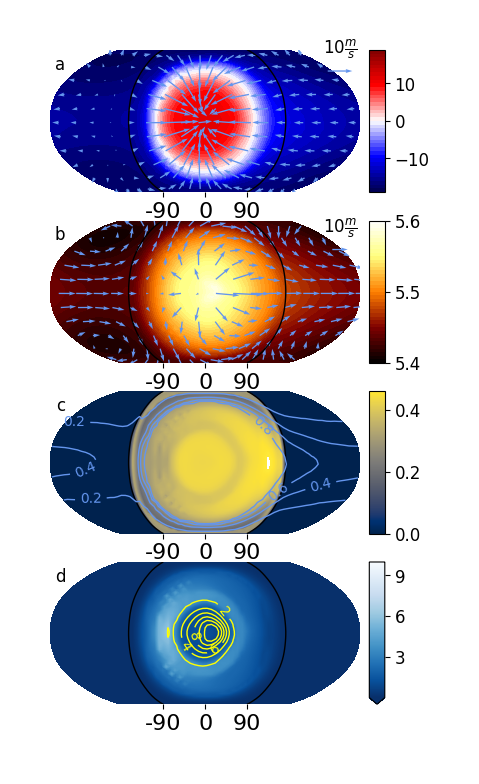}
\caption{\label{fig:wind} a) Near surface ($\sigma = 0.9$) temperatures (\degree C) and wind patterns (quivers). b) Mid-troposphere ($\sigma = 0.5$) geopotential height (km) and wind patterns. c) TOA albedo and vertically integrated total cloud fraction (blue contours), d) Surface evaporation rates and precipitation reaching the surface in yellow contours ($10^{5}$  kg/m$^2$s)}.
\end{figure}

Increasing the mean planetary temperature can result in many changes to the radiative budget, including competing effects. For example, increasing water vapor contributes to cloud formation and an increase in planetary albedo, which has a strong effect on the climate of M-dwarf planets \citep{Yang2013, Yang2014lowordermodel}. As can be noted in Fig.~\ref{fig:wind}c, regions with abundant clouds reflect a larger fraction of incoming SW, which tends to reduce warming particularly in the substellar region. Meanwhile, increased water vapor also leads to stronger LW absorption and a stronger greenhouse effect. But, individually, neither of these effects supports a reduction in the day-night temperature contrast. While, the presence of clouds would reduce dayside warming, the cloud fraction is already high in Aq34, such that there is not a significant difference between the top-of-atmosphere (TOA) albedo in Aq34 and Aq30. Meanwhile, the increased dayside greenhouse effect on its own, in the absence of any energy and moisture transport, would actually promote increased day-night temperature contrasts. Clearly that is not what occurs, as can be noted in the surface temperatures (Fig. \ref{fig:min_max_T}a) and precipitable water (Fig. \ref{fig:min_max_T}b). To fully understand the relationship between the day and nightside climate, we have to explore the relationship between the radiative budget and the atmospheric energy transport. 

\begin{figure}[h]
\centering
\includegraphics[width=.5\textwidth]{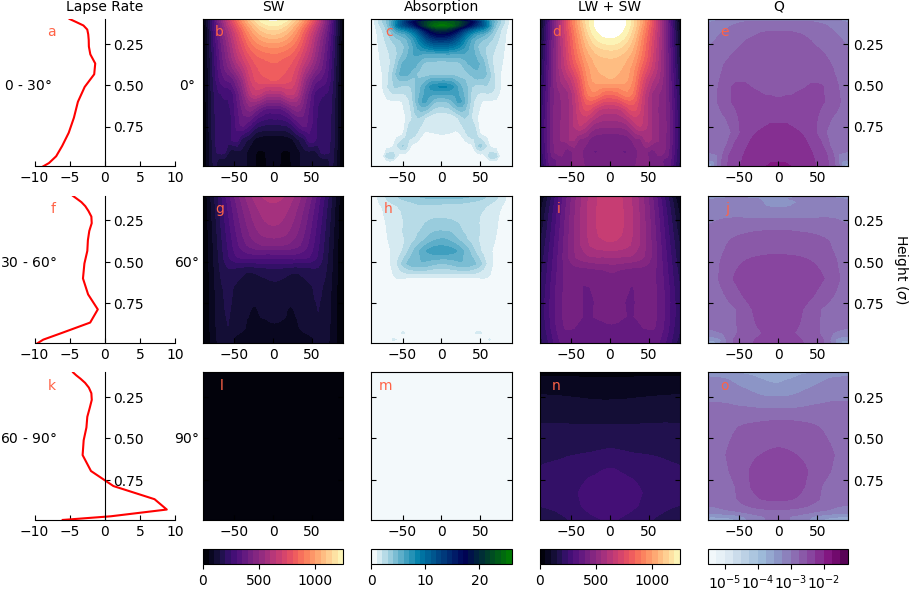}
\caption{\label{fig:sw_and_q} Radiative profiles for Aq34. Left column shows temperature lapse rate (K/km) calculated with area weighted average of region a: 0-30\degree, f: 30-60\degree, k: 60-90\degree \space from substellar point. The remaining columns, from left to right, show downward SW (W/m$^2$); solar heating rate (K/day); Total downward radiation (SW + LW); Specific Humidity (kg/kg). The cross sections were taken at longitudes 0\degree, 60\degree, and 90\degree. }
\end{figure}

As can be observed for Aq34, in Fig.~\ref{fig:sw_and_q} (e,j,o), water vapor (Q) maximizes near the substellar region, and has large values even in the mid-troposphere. The water vapor leads to peaks in SW absorption in the mid and upper atmospheric layers (Fig.~\ref{fig:sw_and_q}c,h), which when combined with reflection from clouds, prevents most of the SW from reaching the surface (b,g,i). The atmospheric substellar absorption of M-dwarf radiation on an Earth-like planet is stronger than for F or G-dwarf instellation \citep{Shields2019}, such that in this simulation less than 50 W/m$^2$ of SW reaches the surface. Even with 400 W/m$^2$ of LW, the substellar surface is a radiative local minimum (Fig.~\ref{fig:sw_and_q},d). Near-surface downwards radiative flux maxima occur in the ``mid-latitudes" (roughly 50\textdegree \space from the substellar point), creating regions of strong evaporation. These are also regions of substantial sensible energy and moisture flux divergence due to the large-scale overturning circulation's near-surface winds patterns, such that the ``mid-latitude" temperatures are more moderate. Meanwhile, the resulting transport of sensible energy and moisture helps sustain the substellar region's temperature maxima, despite the weak radiative values.

\begin{figure}[h]
\centering
\includegraphics[width=0.5\textwidth]{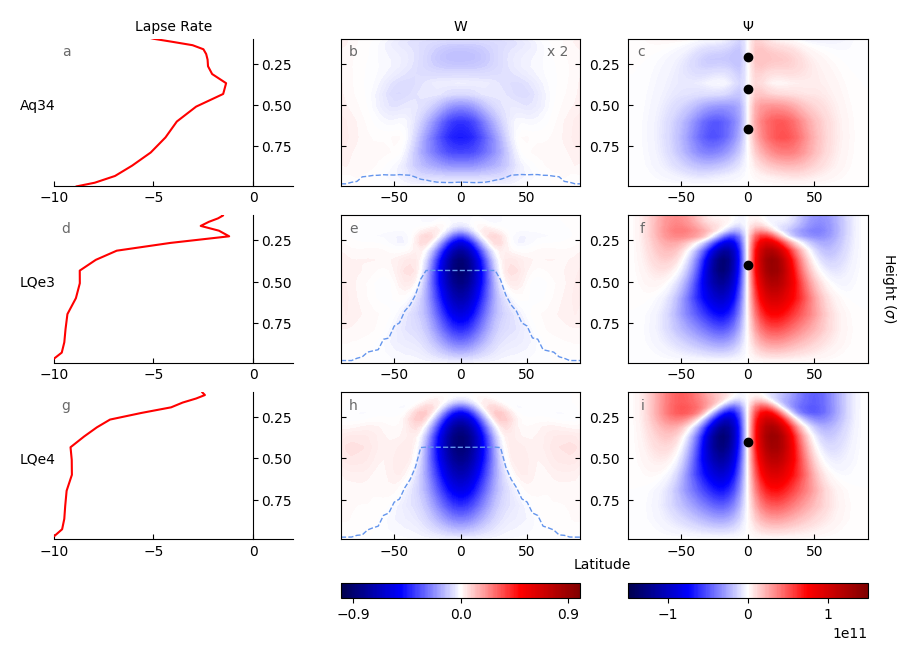}
\caption{\label{fig:lapse_w_strm} Comparison of aquaplanet (Aq34) and land planet (L34Qe3 and L34Qe4) simulations with orbital period of 34 days. Left column shows atmospheric lapse rate, averaged between 0\degree and 30\degree \space from the substellar point. The middle column shows vertical velocity (Pa/s), and the boundary layer depth (dashed line). The right column shows mass stream functions (kg/s), and black markers indicate levels where substellar horizontal flow divergence switches sign. The velocity and stream function cross sections are averaged zonally from -15\degree \space to 15\degree \space latitude. Vertical velocity values for panel b are multiplied by a factor of 2 to allow for easier comparison.  }
\end{figure}

With little SW reaching the surface, and most stellar radiation being absorbed in the mid and upper atmosphere, we would expect weak atmospheric lapse rates in the aquaplanet simulations. Fig.~\ref{fig:sw_and_q}a shows atmospheric lapse rates for Aq34, averaged within 30\degree \space radius of the substellar point. The near-surface lapse is roughly adiabatic (8.8 K/km), and decreases above the boundary layer in the region of strong convection, as would be expected for a region with moist convection. In the region near $\sigma=0.5$, where we note strong SW absorption (Fig.~\ref{fig:sw_and_q}c), the lapse rate decreases further, dropping below 4 K/km at $\sigma = 0.4$. This is significantly lower than the local saturated adiabatic lapse rate of $\sim$7 K/km, creating a region of convective stability that limits the vertical flow (Fig.~\ref{fig:lapse_w_strm}b).

Overall, the atmospheric circulation follows a structure typical for a planet in a slowly rotating regime \citep[e.g.][]{Merlis2010, Showman2013a, HaqqMisra2018}. There is strong near-surface convergence at the substellar region (Fig.~\ref{fig:wind}a), which drives an upward flow. The flow then diverges in the mid/upper troposphere (Fig.~\ref{fig:wind}b), and sinks in the nightside of the planet. But, due to the region of convective stability, the overturning circulation structure is shallow (\ref{fig:lapse_w_strm}c) and above the region of convective stability, a second smaller cell forms. The divisions between the cell layers are indicated with black markers in Fig.~\ref{fig:lapse_w_strm}c. If we considered only the cell's appearance, and the relatively small mass transport, we might erroneously assume that atmospheric transport plays a reduced role in the water-abundant simulations.

While the atmospheric circulation intensity reflects wind speeds and mass transport, it does not necessarily provide a good indication of the net energy transport. For a planet at steady state, in the absence of ocean energy transport, any TOA imbalance must be equivalent to the vertically integrated atmospheric energy transport, such that the energy budget \citep[e.g.][]{Neelin2007} for a synchronously rotating planet without a seasonal cycle, can be simplified to:

\begin{equation} \label{eq: energy_vint}
     R_{\text{toa}} - F_{\text{sfc}} = \nabla \cdot \langle \overline{\mathbf{v}h} \rangle,
\end{equation}
 where $R_{\text{toa}}$ is top-of-atmosphere radiative fluxes, and $F_{\text{sfc}}$ the surface fluxes, which include radiative, sensible, and latent heat flux at the surface. The moist static energy term is defined such that $h = c_{p} T + L_v Q + g Z $, comprised of dry enthalpy $c_{p} T$, latent energy $L_v Q$, and potential energy $gZ$. $\langle \,\cdot \, \rangle$ denotes a vertical integral, and $\overline{(\, \cdot \,)}$ denotes a temporal long-term average.

\begin{figure}[h]
    \centering
    \includegraphics[width=0.4\textwidth]{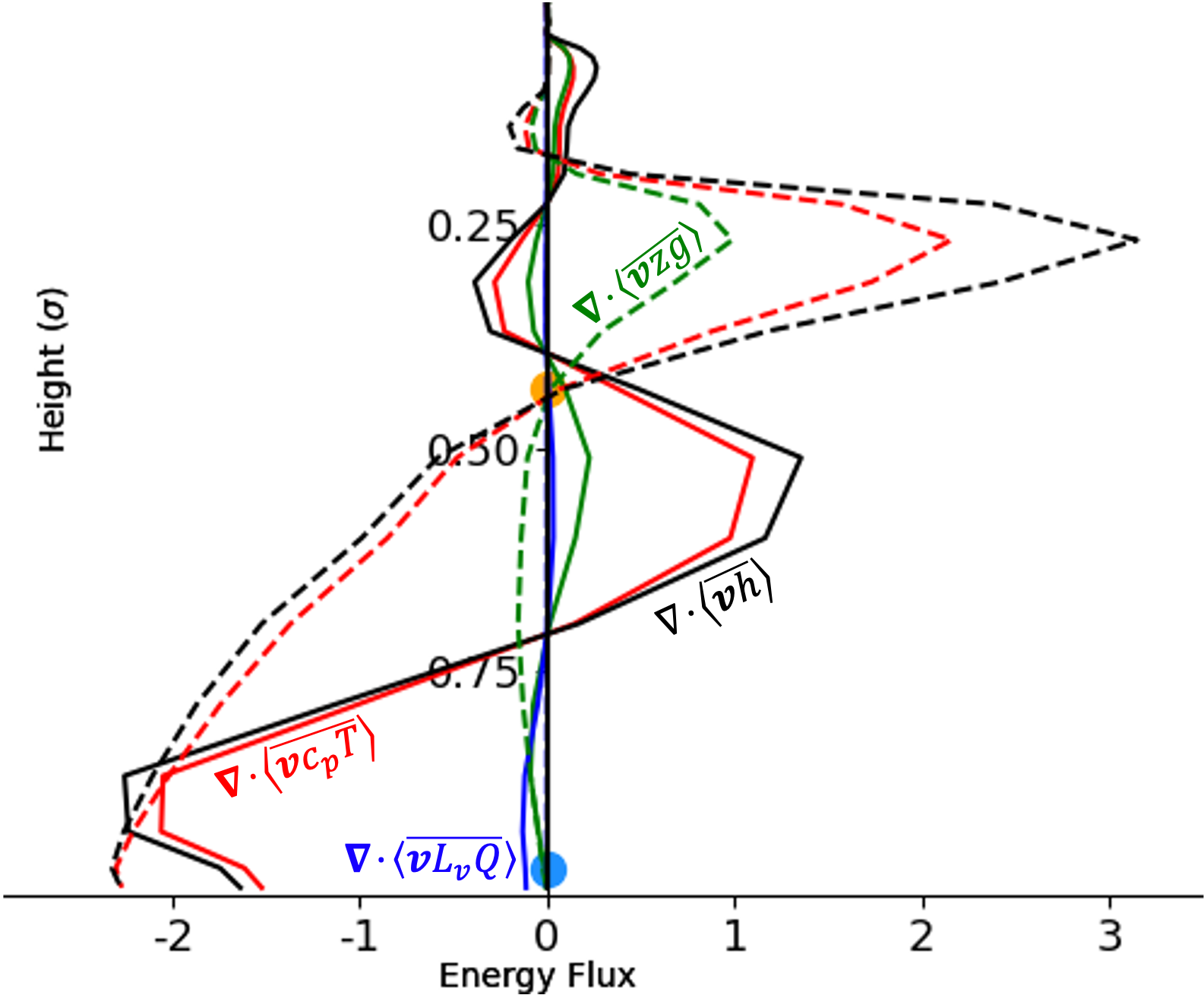}
    \caption{Energy flux divergence in substellar region, for Aq34 (solid lines) and L34Qe4 (dashed lines). Energy fluxes are broken down into sensible (red), latent (blue), and potential energy (green) components, averaged across area within 30\textdegree \space latitude of the substellar point, and weighted by vertical density profile to facilitate comparison of upper and lower level energy transport. Markers show the substellar boundary layer height for Aq34 (blue) and L34Qe4 (orange).}
    \label{fig:vertical_divergence}
\end{figure}

\begin{figure}[h]
\centering
\includegraphics[width=0.49\textwidth]{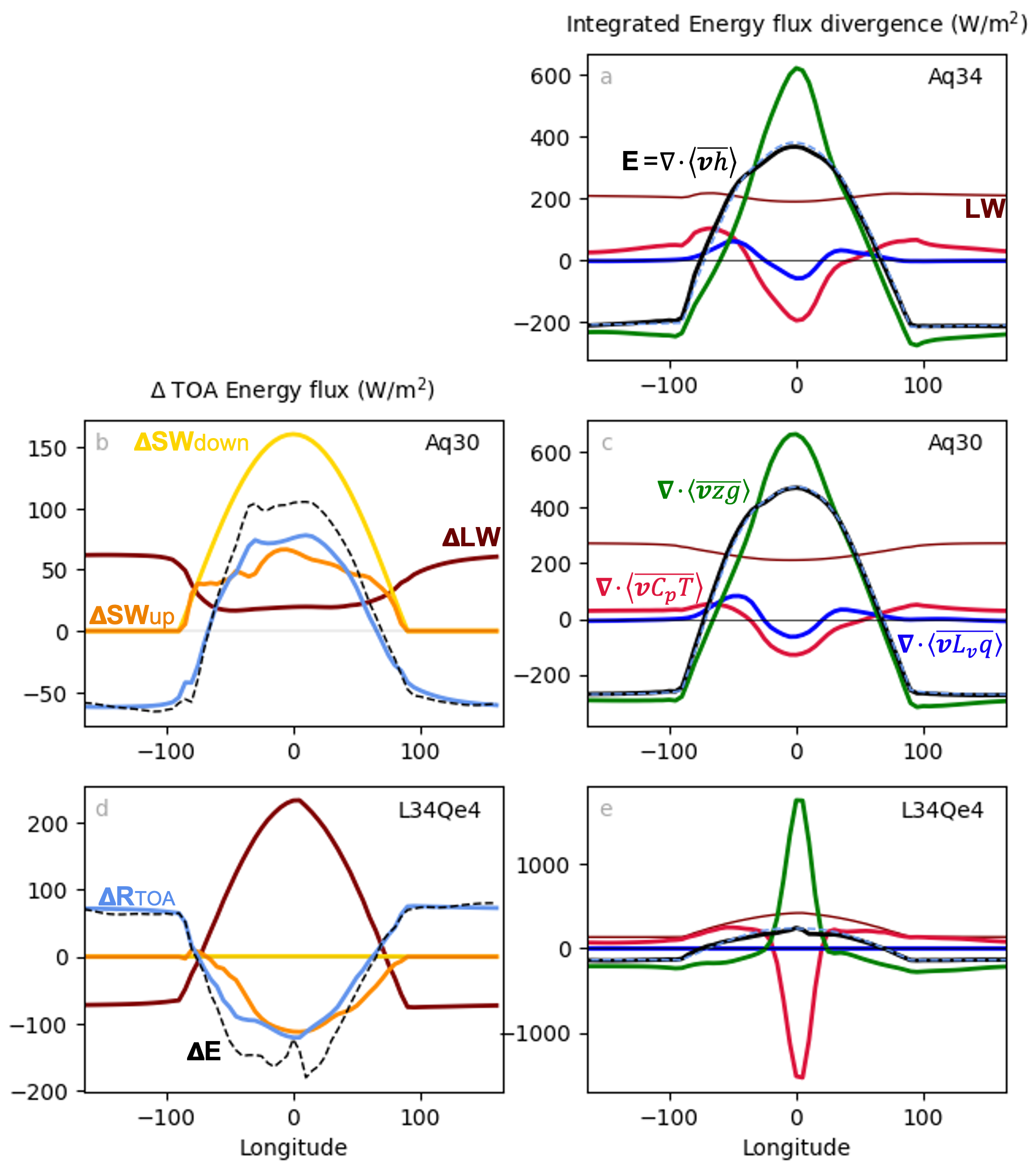}
\caption{\label{fig:NEI} Breakdown of the planetary energy budget. The left column shows changes in the global mean radiative budget relative to Aq34, such that $\Delta X = X - X_{Aq34}$, where X is incoming stellar radiation $SW_{down}$ (yellow), TOA reflected SW ($SW_{up}$, orange), TOA outgoing LW (maroon), the net TOA radiation ($R_{TOA}$, in light blue), and the net energy flux divergence ($E = \nabla \cdot \langle \overline{\mathbf{v}h} \rangle$, dashed black line). For each simulation, the actual energy flux divergence values are plotted on the right, broken down into sensible (red), latent (blue), and potential energy (green) components, vertically integrated and meridionally averaged. The thin maroon line shows TOA upward LW values. We also plot net atmospheric energy input ($R_{\text{toa}} - F_{\text{sfc}}$), as a dashed light blue line, which overlaps with the net energy flux divergence, confirming a closed budget.}
\end{figure}

As is the case with Earth's mean meridional circulation, the near-surface convergence of warm air (in this case at the substellar point) results in net convergence of dry enthalpy (or sensible energy), and latent energy. The higher altitude divergent flow (Fig.~\ref{fig:wind}b), tends to transport colder and dryer air, but with larger amounts of potential energy. On Aq34, both branches transport a significant amount of sensible energy (Fig. \ref{fig:vertical_divergence}), such that the sensible energy transport partially cancels out when vertically integrated. The latent energy transport is comparatively small, such that the potential energy transport of the upper branch ultimately determines the sign of the vertically integrated energy transport (Fig.~\ref{fig:NEI}a). The small value of the latent energy flux divergence indicates that, while water clearly impacts the overall climate outcome, the latent energy transport has only a small effect on the temperature gradients. 
The energy flux divergence (positive values in Fig.~\ref{fig:vertical_divergence} and ~\ref{fig:NEI}) indicates net energy transport away from the substellar point. 

As we increase the stellar flux (Aq30), the increased dayside SW ($\Delta SW_{down}$), is slightly reduced due to reflection from the TOA albedo ($\Delta SW_{up}$), and balanced partially by increased LW ($\Delta LW$), and mostly by increased energy transport divergence ($\Delta E$). This can be noted in Fig.~\ref{fig:NEI}b, which compares Aq30 to Aq34. The net energy transport remains similar in shape (Fig.~\ref{fig:NEI}a,c), but the net sensible energy transport is slightly reduced in favor of latent and potential energy transport. 

Many of the trends noted for Aq30 persist for Aq25. Simulations that do not achieve steady state, as is the case for Aq25, must be interpreted with caution. Prior to reaching numerical instability, day-to-night temperature differences become nearly negligible (Fig.~\ref{fig:min_max_T}a). The energy budget is not properly closed, as we might expect for a simulation not at steady state. But we notice sensible energy transport becomes small, likely due to the reduced temperature gradients, and the majority of the transport is done by potential and latent energy transport, which increase to similar magnitudes.  

In summary, we can attribute the reduction of day-night temperature differences in our water abundant planets to an increase in net energy transport, which diverges additional energy away from the dayside, and converges additional energy in the nightside. In other words, for increased stellar radiation, the transport increases, reducing the intensity of dayside warming, and enhancing nightside warming. With higher surface and atmospheric temperatures, as well as increased moisture flux convergence, nightside atmospheric water vapor increases (Fig. \ref{fig:min_max_T}b), allowing for additional nightside warming due to an increase in the local greenhouse effect. 

As previously mentioned, the atmospheric energy transport is responsible for resolving radiative imbalances (Eq. \ref{eq: energy_vint}), such that these changes in the atmospheric transport can also be more simply interpreted in the context of the top-of-atmosphere radiative budget. As described in \citet{Koll2016}, for planets in a weak-temperature-gradient regime, the circulation can be thought of as a heat engine driven by dayside heating, and cooling due to global thermal emissions to space. 
The aquaplanet simulations have high atmospheric water vapor content, resulting in optically thick dayside atmospheres, such that their dayside thermal emissions are determined by the upper tropospheric temperature. In such cases, increasing the stellar flux has a small impact on the outgoing longwave radiation (OLR) \citep{Yang2014lowordermodel}, as can be noted in Fig. \ref{fig:min_max_T}c and Fig. \ref{fig:NEI}b. Meanwhile, the colder nightside has low humidity and a relatively transparent atmosphere (except for Aq25), such that warming is closely tied to increased OLR. Therefore, in a moist and opaque atmosphere, increasing the stellar radiation as we go from Aq34 to Aq30 results in a small dayside OLR increase, a larger nightside OLR increase, and requires a corresponding increase in day-to-night energy transport. The extra SW absorbed in Aq30 is primarily balanced by a larger nightside OLR, which is only achievable through a strengthening of the atmospheric energy transport and an overall reduction in day-to-night temperature gradients.

Given that atmospheric energy transport is essential for reducing the day-to-night temperature contrast, we might expect to see larger gradients on planets with thinner atmospheres. However, that is not observed in our aquaplanet simulations where we half the background N$_2$ values. As can be noted in Fig.~\ref{fig:min_max_T}a and Fig.~\ref{fig:TS_robinson}, the temperature range remains almost unchanged when the surface atmospheric pressure is halved. Instead, we note overall increased wind speeds, such that the energy transport terms remain roughly similar. These results are in general agreement with the findings from \citet{Zhang2020}, which showed that background gas pressure has a relatively small impact on the inner edge of the habitable zone for synchronously rotating Earth-like planets. They show that the effects are non-monotonic, and significantly less important than changes in the cloud scheme \citep{Bin2018} and rotation rate \citep{Kopparapu2017}. Therefore, we would not expect that pressure alone would determine the viability of terminator habitability for the Earth-like planets we are considering.

\subsection{Water-Limited Planets} \label{sec:water limited}

Here we switch our focus to water-limited land planets, using ExoCAM with a land planet setup, initialized with varying amounts of atmospheric water vapor (L34Qe3 and L34Qe4). Once temperatures equilibrate, these simulations have 20\% and 1\% of Earth's atmospheric precipitable water respectively, with small amounts of water also in the form of surface ice and soil water, concentrated on the nightside. We will begin by considering the differences between Aq34 and L34Qe4.

With reduced atmospheric water vapor, we expect a reduction in cloud coverage, especially for low-level clouds. This expectation is confirmed, and results in a reduction of the dayside albedo, which becomes largely homogeneous with values between 0.2 and 0.26 in the water-limited simulations. There is still some condensation, precipitation, and high altitude cloud formation, occurring near the substellar point. But, with re-evaporation in the atmosphere, surface precipitation becomes negligible. These precipitation patterns, noted in all of our land planet simulations, appear to follow the ``transition regime" described in \citet{Ding2021}. Despite having the same stellar flux as Aq34, L34Qe4 has a larger net SW due to reduced cloud coverage and lower albedo (negative $\Delta SW_{up}$ in Fig.~\ref{fig:NEI}d). The resulting increase in substellar net SW must result in one of two outcomes: either increased TOA LW emissions, or increased day-to-night energy transport. 

\begin{figure}[h]
\centering
\includegraphics[width=0.5\textwidth]{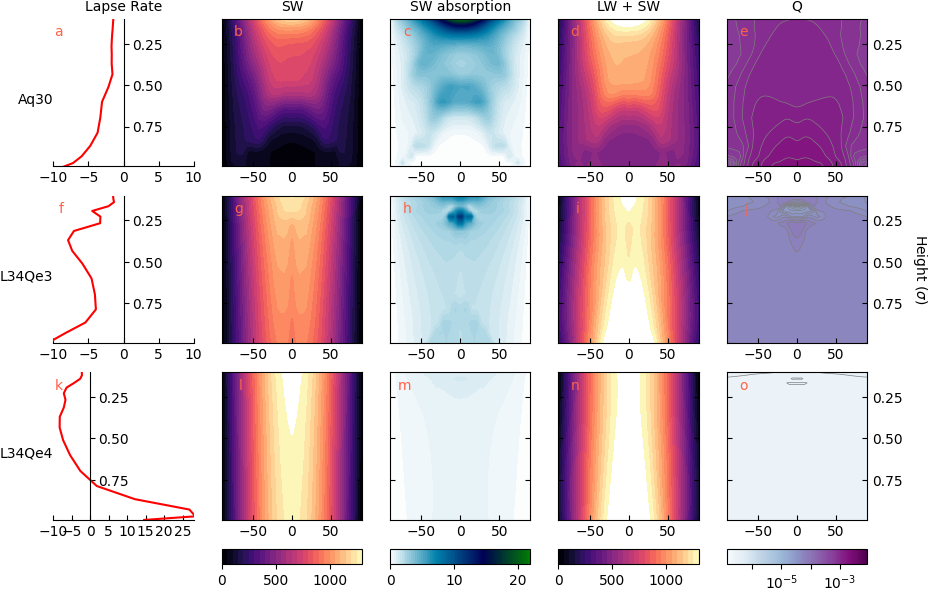}
\caption{\label{fig:sw_and_q_allsim} Same as top row of Fig.~\ref{fig:sw_and_q}, with each row showing lapse rates averaged over a region 0-30\degree \space from the substellar point and cross sections taken at 0\degree \space longitude, for Aq30, L34Qe3, and L34Qe4. To facilitate comparison across simulations, specific humidity is shown on a logarithmic scale.}
\end{figure}

Near the substellar point, a much larger fraction of SW reaches the surface (Fig.~\ref{fig:sw_and_q_allsim}), with dayside surface temperatures reaching maxima of 355K (L34Qe4). Though there is still water-vapor SW absorption, the atmosphere's vertical temperature structure more closely resembles a dry region on Earth, with dayside lapse rates remaining steeper than 6K/km until $\sigma = 0.25$ (Fig.~\ref{fig:lapse_w_strm}). Without the low region of convective stability noted in Aq34, the water-limited simulations have deeper overturning cells, ascending at the substellar point (Fig.~\ref{fig:lapse_w_strm},f,i). 

Examining the energy flux divergence (Fig.~\ref{fig:NEI}e), we note a strengthening of the individual energy transport terms, but a weakening of the net transport. The deepening of the atmospheric circulation, combined with strong high altitude winds, leads to significantly larger values of potential energy transport away from the ``eye", as can be noted in Fig. \ref{fig:vertical_divergence} which shows the vertical profile of the energy flux divergence averaged across an area within 30\textdegree \space of the substellar point. Sensible energy fluxes are intensified in both the upper and lower branches of the circulation, but the lower branch increases significantly in both intensity and depth. The boundary layer in the water-limited simulation is significantly deeper (Fig. ~\ref{fig:vertical_divergence}), and the lower branch of the circulation occupies its full depth, reaching up to $\sigma = 0.4$. This effect can also be noted in the upward shift of the cell's core (Fig.~\ref{fig:lapse_w_strm}). This ``bottom-heavy" cell structure achieves higher values of sensible heat conversion at the substellar point, which largely cancels out the increased potential energy divergence, such that the net day-to-night energy transport is reduced relative to Aq34 despite the increased net SW. Comparing $\Delta SW_{up}$ to the change in net energy flux divergence ($\Delta E$), we note that the effects of the reduced atmospheric energy transport are of comparable intensity to the effects of the reduced albedo. 
 
With reduced albedo and reduced energy transport, the large TOA imbalance is resolved by an increase in dayside TOA LW emissions (positive $\Delta LW$ Fig.~\ref{fig:NEI}d). It is interesting to note that in the water-limited cases, the substellar region becomes a region of LW maxima, rather than a minima as was the case for the aquaplanet simulations (Aq34 and Aq30). Meanwhile, the reduced nightside energy convergence results in a reduction of nightside temperatures. Together, these effects allow the water-limited case to achieve equilibrium with larger day-night temperature gradients, spanning 139K (L34Qe4) and 122K (L34Qe3). If we assume that surface temperatures surpassing 50\degree C are hostile to life, and temperatures nearing 100\degree C would be uninhabitable, both L34Qe3 and L34Qe4 water-limited scenarios can be considered cases of terminator habitability. 

As with the aquaplanets, we can re-frame these results in terms of the atmospheric radiative imbalances to obtain a simpler intuition. On these land planets, the atmosphere is dryer, and therefore less cloudy and opaque. This allows the dayside to be closer to radiative equilibrium (Fig. \ref{fig:NEI}d), and therefore implies a reduced energy transport to the nightside. The water-limited L34Qe4 absorbs a greater fraction of the incoming stellar radiation at the surface than Aq34, achieving higher surface temperatures, and higher OLR, thus requiring a weaker energy transport (F ig. \ref{fig:NEI}d). In turn, the nightside receives less energy, and is colder than the aquaplanet counterpart. This results in significantly larger day-night temperature contrasts, such that a temperate terminator becomes not only easily achievable, but also a likely scenario for land planets in the habitable zone. 

The simulations discussed so far illustrate the possibility of terminator habitability as an alternate climate configuration within the habitable zone (at orbital period of 34 days). However, it is also worth emphasizing that water-limited planets can endure higher stellar fluxes without entering a runaway greenhouse state. For example, while Aq25 and Aq25h become unstable due to runaway warming, an equivalent water-limited simulation (L25Qe4) achieves steady state and sustains a temperate terminator climate (Fig.~\ref{fig:TS_robinson}f). The increase in substellar SW relative to L34Qe4 is balanced in roughly equal parts by an increase in LW and an increase in local energy flux divergence. The resulting climate is on average warmer than L34Qe4, and the day-to-night temperature range increases to 154K (Fig.~\ref{fig:min_max_T}). The L25Qe4 nightside temperatures remain just below freezing, dayside temperatures reach scorching highs over 400K, and the terminator remains within the 270-320K range. While the fractional habitability (based on surface temperatures) of L25Qe4 may be smaller than Aq34 and Aq30 (Table~\ref{tab:mean_values}), the contrasting results of L25Qe4 and Aq25 illustrate how we might expect to observe planets with diverse surface climates and regions of temperate surface temperatures even at the inner edge of the habitable zone.

\subsection{Long-Term Stability} \label{sec:long term}

The water-limited simulations show that terminator habitability is possible. However, based on the results discussed so far, we have not yet established whether they are ideal observational target in the search for life beyond Earth. After all, these simulations depict planets with harsh climates, limited regions of habitability, and limited water availability, which may seem unappealing when compared to the climates of ocean worlds obtained with aquaplanet simulations. However, arid exoplanets offer observational advantages, because their reduced cloud coverage could facilitate detection of water \citep{Ding2022} by JWST, potentially making them more practical near-future targets. Also, 
a closer look at the aquaplanet moisture budget shows that we must be cautious of assuming long-term climate stability, which is crucial for increasing the odds of developing life. 

Aquaplanet simulations provide the atmosphere with an unlimited water source. On the planet's dayside, not only is there water available everywhere, but there is no depletion in regions of negative net precipitation, nor resistance to evaporation as would occur from drying soil \citep[e.g.][]{Griend1994}. For Aq34, which has dayside temperatures comparable to Earth's, this results in an Earth-like amount of atmospheric precipitable water ($\sim$70\% of Earth's atmospheric water). The ``eye" is a region of negative net precipitation, and there is continuous transport of water vapor to the nightside, resulting in a total of 1.5 x 10$^{16}$ kg/s of snowfall. For scale, this rate of snowfall would cold-trap a volume of water equivalent to Earth's oceans in 90 thousand years. The presence of a deep global ocean circulation could, in some cases, help deglaciate part or all of the nightside ocean \citep{Hu2014}. But in the absence of a global ocean with favorable properties, such as large depth \citep{Hu2014} and high salinity \citep{Olson2020}, we can assume a limited return flow for water deposited on the nightside. Even accounting for ice sheet dynamics, which would imply a down-gradient flow of ice towards the dayside, it is possible that over time a majority of the water from an ocean-covered Earth-like planet could become cold-trapped \citep{Menou2013, Leconte2013}. 

While Aq34 could be at risk from cold-trapping, the risk decreases significantly for planets closer to the star. The Aq30 nightside temperatures are higher, such that the equatorial region of the nightside remains above freezing. The snowfall is confined to the higher latitudes, and the global snowfall rate decreases by a factor of 100 relative to Aq34. However, the atmospheric precipitable water increases to approximately 3.5 times Earth's atmosphere. In particular, we note an increase in water vapor content above the atmospheric cold trap, reaching values of 0.003 kg/kg ($\sim$0.005 mol/mol), which is two orders of magnitude larger than Aq34 values, and reaches the limit proposed by \citet{Kasting1984} and confirmed in a series of planetary studies \citep[e.g.][]{Kasting2015, Wolf2015}. This raises the possibility that such a planet would be vulnerable to continuous water vapor loss, especially during periods of strong flare activity. These losses are in addition to water losses resulting from enhanced heating during the M-dwarf's protracted pre-main sequence phase, which could impact all of these scenarios and cause the loss of large water reservoirs \citep[e.g.][]{Bolmont2017, Luger2015}. Therefore, we have abundant reason to question water-rich M-dwarf scenarios.

Meanwhile for land planets L34Qe3, L34Qe4, and even L25Qe4, while there is still risk of water loss in their early history, high altitude moisture values are comparable to Aq34 and well below the \citet{Kasting1984} limit. Similarly, despite their extremely low nightside temperatures, at equilibrium there is no snow being deposited on the nightside. There is some snow and ice accumulation in L34Qe3 while the model spins up, which remain trapped. But the remaining moisture is subsequently retained in the atmosphere above the planetary boundary layer, primarily on the dayside. We have confirmed that simulations initialized with more water (e.g. Q = 0.01) have similar night-side water trapping, and slowly tend toward the climate in L34Qe3, but would require hundreds more simulated years to fully reach equilibrium. 

Based on our land planet simulations, water-limited atmospheres appear to reach a more stable configuration than their aquaplanet counterparts, being less vulnerable to additional water vapor loss or nightside cold-trapping. It also suggests that ocean world configurations subject to significant water loss, without entering a runaway greenhouse state, could eventually reach a stable configuration with terminator habitability. These considerations, combined with the observational advantages of water-limited planets \citep{Ding2022}, suggests that despite their potentially limited or localized fractional habitability, land planets will be important observational targets in the coming years, and will play a prominent role in the early stages of exoplanet climate characterization.

\subsection{Terminator Water Availability} \label{sec:water_avail}

Thus far we have defined terminator habitability in terms of surface temperature. However, water is also a crucial ingredient for life as we know it, and could pose a problem for water-limited scenarios. In our simulations, rain is strongly concentrated in the substellar region (for both aquaplanet and land-planet simulations), and snow occurs across the nightside in some Aq simulations. For the rotational configurations considered here, the terminator never receives significant amounts of precipitation. Therefore, even though terminator evaporation rates tend to be low, in our simulations it is a region of weak negative net precipitation. In other words, based on the atmospheric moisture budget, the terminator is a region that would tend to have an arid climate. However, water availability on land planets is not exclusively determined by the atmospheric moisture budget, but also depends on factors such as groundwater transport and glacier behavior. These effects are in turn dependent on many poorly constrained planetary properties such as orography, soil structure, and geothermal heat flux. While we cannot quantify these effects with our climate model, we can discuss them qualitatively to explore how they would impact terminator water availability. 

The role of glacier flow has been widely discussed both in the context of snowball Earth \citep[e.g.][]{Goodman2003, Pollard2005} and as a potential mechanism for releasing cold-trapped water on synchronously rotating planets \citep[e.g.][]{Leconte2013}. Ice sheets can deform under their own weight, resulting in an ice flow down the ice thickness gradient. On a synchronously rotating planet, this flow could push ice towards the terminator, where the temperate climate would result in melting. The flow velocity depends on, among other things, the ice geometry, planetary gravity, and geothermal heat flux. For Super-Earths, \citet{Yang2014WaterTrapSuperEarth} argue it could largely prevent nightside cold-trapping on water abundant planets. 

Generally, the ice flow would be slower on Earth-sized planets, especially those with more moderate geothermal heat fluxes, but it could still provide a large enough source of melt-water to create a moist climate in the terminator region. For example, on Greenland and Antarctica flow speeds and discharge into the ocean can reach rates over 1km/year \citep{Rignot2011, Rignot2012}. Currently, model representations of glacier dynamics are limited, and not fully suitable for planetary purposes. For CESM, even a configuration with an evolving ice sheet, where glacier area and elevation are adjusted so as to conserve mass and energy, would not capture the dynamics relevant for the ice flow we consider here. But, if we make a conservative estimate, considering only the slower flow rates observed at the Antarctic glacier edges ($\sim 100$m/yr), and assuming only a shallow ice layer reaches the terminator ($\leq$ 10m), that would imply water transport rates of order 10$^{-2}$ kg/s. This exceeds the terminator evaporation rates in the Aq30 and Aq34 simulations (see Fig. \ref{fig:wind}d) by two orders of magnitude. Given that aquaplanet simulations have infinite water availability, we can cautiously treat their evaporation rates as an upper limit, such that the comparison of glacier flow to Aq terminator evaporation rates indicates that glacier flow could sustain surface water near the terminator. Depending on the exact rates, this could result in swamp-like regions, oceans, or in smaller lakes and rivers in regions of topographic lows. Even if flow rates were significantly lower than expected and surface runoff resulted in a large evaporating area, soil evaporative resistance would tend to retain some local moisture that could be beneficial for the development of life in the terminator region. 

Of course, not all planets would have nightside glaciers. The presence of large ice deposits presumes a planet with abundant initial water, that was either retained throughout the early period of intense stellar emissions, or delivered later on. It also presumes that ice would be stable on the nightside. For planets receiving sufficiently high instellation, the nightside ice can sublimate \citep{Ding2020}, such that water would likely only be present in vapor form. Our arid land planet simulations are not ideal for exploring this limit, but based on the values from \citet{Ding2020}, we would expect nightside sublimation to be important for planets in L25Qe4's orbit. Therefore, L25 planets would be more likely to have a temperate but dry terminator than their L30 or L34 counterparts. 

A less recognized, but also potentially important mechanism to consider for planets with stable surface water, is groundwater transport. \citet{Faulk2020} showed that surface and subsurface methane flows on Titan could help shape the global hydrological cycle, transporting liquid from the low-latitude highlands to the lower polar regions, resulting in a moist polar climate. For synchronously rotating planets, \citet{Turbet2016} argues alignment of large-scale gravitational anomalies and the star-planet axis is favored \citep{Wieczorek2007}, such that the largest topographic basins would tend to be near the substellar point or anti-stellar point, which has served as motivation for studies of sustellar continents on ocean worlds \citep{Lewis2018}. For a land planet, where precipitation is strongly concentrated at the substellar region, a substellar topographic high could result in a hydrological cycle that continuously transports water across a larger portion of the dayside. Whereas a topographic low could further concentrate water near the substellar point. Therefore, additional exploration of topography, runoff and subsurface transport is necessary to constrain surface water availability for a land planet, especially for those with significant large-scale topography.

Based on the above discussion of water availability, we might expect that some, but not all planets with temperate terminators would also have moist terminator climates. We expect that future work, exploring a wider range of surface configurations will help better constrain water availability for land planets. Model advancements to include glacier dynamics into the water budget would further improve our understanding of these water worlds. Finally, an important caveat to this work is the fact that we have only considered planets with an Earth-like atmospheric composition. Varying atmospheric mass and composition would certainly impact these results. For example, \citet{Ding2020} showed that nightside cold-trapping might be avoided on planets with high CO$_2$ values, due to increased nightside temperatures. Constraining exoplanet CO$_2$ levels remains a challenge, due to their dependence on outgassing and weathering rates \citep{Walker1981}. Therefore, future work may wish to consider both terminator habitability and high CO$_2$ ``eye" habitability scenarios, while exploring additional factors such as the impact of groundwater flow and long term climate sensitivity in the presence of M-dwarf flare activity to quantify prospects for sustained habitability.

\section{Conclusions} \label{sec:Conclusions}

In this study we explored the possibility of terminator habitability, defined as the existence of a habitable band at the transition between a scorching dayside and a glacial nightside on a synchronously rotating Earth-like planet orbiting an M dwarf. In particular, we explore the viability of this configuration at the inner edge of the habitable zone to determine if it is possible for dayside temperatures to exceed typical habitable temperatures, approaching the runaway greenhouse limit, without leading to a planet-wide runaway greenhouse state. 

We find that a temperate terminator is not achievable with aquaplanet simulations that seek to reproduce ocean-covered planets, but can easily occur on water-limited land planets. On aquaplanets, increasing stellar flux leads to reduced day-night temperature gradients, such that the planet would tend to reach a homogeneous climate before the dayside reaches a runaway greenhouse state, never passing through a terminator habitability state. Whereas on water-limited planets, we find that large day-night temperature gradients are easily achievable without entering a runaway greenhouse state. We also find that the water-limited land planet configurations may be favorable in terms of long term climate stability, with reduced risk of nightside water cold-trapping or water vapor escape. We expect that water-limited terminator habitability scenarios could be a stable configuration for ocean-covered worlds after significant water loss, especially in the case of nightside cold-trapping. 

There are still many uncertainties regarding the water content of habitable zone M-dwarf planets. Based on our current understanding, it is possible that water-limited planets could be abundant and possibly more common than ocean-covered worlds. Therefore, terminator habitability may represent a significant fraction of habitable M-dwarf planets. Compared to the temperate climates obtained with aquaplanets, terminator habitability does offer reduced fractional habitability. Also, while achieving a temperate terminator is relatively easy on water-limited planets, constraining water availability at the terminator remains a challenge. Overall, the lack of abundant surface water in these simulations could pose a challenge for life to arise under these conditions, but mechanisms including glacier flow could allow for sufficient surface water accumulation to sustain locally moist and temperate climates at or near the terminator. We expect that future studies exploring a broader range of land planet configurations, in particular those using future generations of surface and ice models, will find a wide range of habitable terminator scenarios in regimes intermediate to the water-limited and aquaplanet cases considered here. 

\section{Acknowledgments}
This material is based upon work supported by the National Science Foundation under Award 1753373, and by a Clare Boothe Luce Professorship supported by the Henry Luce Foundation. This research was also performed as part of the NASA's Virtual Planetary Laboratory, supported by the National Aeronautics and Space Administration through the NASA Astrobiology Institute under solicitation NNH12ZDA002C and Cooperative Agreement Number NNA13AA93A, and by the NASA Astrobiology Program under grant 80NSSC18K0829 as part of the Nexus for Exoplanet System Science (NExSS) research coordination network.
We would like to acknowledge high-performance computing support from Cheyenne (doi:10.5065/D6RX99HX) provided by NCAR's Computational and Information Systems Laboratory, sponsored by the National Science Foundation.

\clearpage

\bibliography{bibs}{}
\bibliographystyle{aasjournal}



\end{document}